\documentclass[sigconf]{acmart}

\usepackage[whole]{bxcjkjatype} % for Japanese on pdfLaTeX
\usepackage[subrefformat=parens]{subcaption}
\usepackage{url}

\AtBeginDocument{%
  \providecommand\BibTeX{{%
    \normalfont B\kern-0.5em{\scshape i\kern-0.25em b}\kern-0.8em\TeX}}}
\copyrightyear{2021} 
\acmYear{2021} 
\setcopyright{acmcopyright}\acmConference[WI-IAT '21]{IEEE/WIC/ACM International Conference on Web Intelligence}{December 14--17, 2021}{ESSENDON, VIC, Australia}
\acmBooktitle{IEEE/WIC/ACM International Conference on Web Intelligence (WI-IAT '21), December 14--17, 2021, ESSENDON, VIC, Australia}
\acmPrice{15.00}
\acmDOI{10.1145/3486622.3493982}
\acmISBN{978-1-4503-9115-3/21/12}

\begin{document}
\title{%
  Do you trust experts on Twitter?}
\subtitle{Successful correction of COVID-19-related misinformation}

\author{Dongwoo Lim}
\affiliation{%
  \institution{Graduate School of Interdisciplinary Information Studies, \\The University of Tokyo}
  \country{Japan}}
\email{ohthekoo@g.ecc.u-tokyo.ac.jp}

\author{Fujio Toriumi}
\affiliation{%
  \institution{Graduate School of Engineering, \\The University of Tokyo}
  \country{Japan}}
\email{tori@sys.t.u-tokyo.ac.jp}

\author{Mitsuo Yoshida}
\affiliation{%
  \institution{Faculty of Business Sciences, University of Tsukuba}
  \country{Japan}}
\email{mitsuo@gssm.otsuka.tsukuba.ac.jp}

\begin{abstract}
This study focuses on how scientifically-correct information is disseminated through social media, and how misinformation can be corrected. We have identified examples on Twitter where scientific terms that have been misused have been rectified and replaced by scientifically-correct terms through the interaction of users. 
The results show that the percentage of correct terms (``variant (変異株)'' or ``COVID-19 variant (変異ウイルス)'') being used instead of the incorrect terms (``strain (変異種)'') on Twitter has already increased since the end of December 2020. This was about a month before the release of an official statement by the Japanese Association for Infectious Diseases regarding the correct terminology, and the use of terms on social media was faster than it was in television. Some Twitter users who quickly started using the correct term were more likely to retweet messages sent by leading influencers on Twitter, rather than messages sent by traditional media or portal sites. However, a few Twitter users continued to use wrong terms even after March 2021, even though the use of the correct terms was widespread. Further analysis of their tweets revealed that they were quoting sources that differed from that of other users. This study empirically verified that self-correction occurs even on Twitter, which is often known as a ``hotbed for spreading rumors.'' The results of this study also suggest that influencers with expertise can influence the direction of public opinion on social media and that the media that users usually cite can also affect the possibility of behavioral changes.
\end{abstract}



\maketitle

\section{Introduction}
Regarding information on COVID-19, which social medium can be trusted? Can Twitter be trusted? Not many people would be happy to say, ``I trust it.'' Twitter is often known as an epicenter of misinformation or rumors. It has been widely accepted that misinformation and rumors spread more rapidly than factual information on social media~\cite{Ratkiewicz:2011}. The dissemination of misinformation can have negative effects on both personal and social levels, especially amid a global pandemic. On a personal level, it has been reported that people face fear, anxiety, and stress when there is a spike in false information related to COVID-19~\cite{Kumar:2021}. The prevalence of misinformation also has a negative impact on society; for example, false information about the relationship between vaccination and autism, and the denial of vaccination as a consequence has caused public health exigencies~\cite{Larson:2018}.
Furthermore, some experts have noted that in a democratic society, decisions made by citizens who are not well informed or are misinformed may be detrimental to the community~\cite{Kuklinski:2000}. In this regard, social media platforms, such as Facebook, Twitter, and Instagram, have also announced that they will actively deal with false information surrounding COVID-19~\cite{Forbes:2020}. Correspondingly, Twitter posted information from the World Health Organization (WHO) or national health authorities about conspiracy theories and misinformation from sites that disseminate false information and used algorithms to identify and eliminate the circulation of potentially damaging false information. 

Meanwhile, we found a notable example of scientific facts dispelling misinformation and becoming the dominant term through spontaneous interactions among Twitter users. The COVID-19 variant that originated around December 2020 was initially called the ``strain (変異種)'' in Japan. However, experts noted that it is appropriate to call it a ``variant (変異株)'' from a scientific point of view. Accordingly, the Japanese Association for Infectious Diseases officially issued a statement on its website on January 22, 2021, directing the Japanese media to use the correct terms ``variant (変異株)'' or ``COVID-19 variant (変異ウイルス).''~\cite{Japanese:2021} Meanwhile, even before this statement was released, there were posts on Twitter in December 2020, stating that ``variant'' was the correct expression, not ``strain.'' In a news program aired in December 2020 on television, Japan's dominant media, both the incorrect and correct terms were used, but the usage of the correct expression seemed to gradually increase. Therefore, we obtained 1) Twitter data and 2) television news programs' metadata and analyzed them closely. Through this analysis, we sought to understand the dissemination of scientific expertise on Twitter. In summary, the three main points we intend to elucidate through this study are:
\begin{itemize}
  \item How has the trend in the proportion of scientifically-incorrect and correct terms used on Twitter changed? How is it different from television news programs?
  \item Which accounts did users who pioneered the use of correct terminology trust as sources of information? Specifically, what are the characteristics of the accounts retweeted by those who show changes in behavior?
  \item What are the characteristics of users who persist in using the wrong term even after the correct terminology has become prevalent? In particular, what sources of information are they referring to?
\end{itemize}

\section{Literature Review}
\subsection{Correction of misinformation }
It is not easy to correct misinformation that has been accepted as truth, because it is stored in memory and affects subsequent judgments, characterized as ``belief persistence''~\cite{Chan:2017, Lewandowsky:2009}. Furthermore, efforts to rectify misinformation may sometimes prove counterproductive, such as making people more dependent on misinformation~\cite{Nyhan:2010}. For example, in Japan, in March 2020, misinformation spread on Twitter that ``tissue papers are out of stock due to the spread of COVID-19.'' Consequently, many stores sold out of tissue papers. According to a study analyzing related tweets, although the disclaimer spread wider than the misinformation, over-purchase of toilet paper continued~\cite{Iizuka:2021}. With these results in mind, we attempt to empirically verify the characteristics of the case covered in our study.

\subsection{Retweets on Twitter}
Retweeting is the key mechanism for information diffusion on Twitter~\cite{Suh:2010}; information is transmitted to other users through a retweet (RT), which has the meaning of social interaction through social media~\cite{Honeycutt:2009}. Users who retweet can be regarded as opinion leaders from the point of view of communication studies that deal with the dissemination of knowledge. Retweets also serve to disseminate information to many people over a short period. A phenomenon in which messages sent from Twitter spread explosively or rank at the top of portal sites is called a ``Twitter bomb.''~\cite{Ratkiewicz:2011} Based on these previous studies, this research interprets retweets as an act of sharing information in an active sense. In addition, by identifying which users were most retweeted by other users, we intend to determine an axis of information distribution.

\subsection{Cross-media usage and news sharing}
Considering the advancement in media channels, we have used multiple media platforms simultaneously, and the manner in which we used them is also diverse. For example, information is disseminated through traditional media such as newspapers and television, or the news shared on social networking sites such as YouTube or Facebook, and often posted on Twitter or Instagram by sharing personal opinions or impressions, along with the sources of news or capture photos. The use of cross-media on Twitter appears in the form of hashtags, retweets, and comments, as well as by sharing hyperlinks. This study focuses on the sharing of hyperlinks, given that the posting of hyperlinks, unlike retweets and comments, is a common practice without users having to follow one another. Hyperlinks are fundamental connective tools that allow users to direct each other in digital spaces while displaying their own interests in specific news and information~\cite{Holton:2014, DeMaeyer:2013, Hsu:2011}. Holton et al.~\cite{Holton:2014} revealed a central social role for hyperlinks, indicating their use in seeking information by soliciting reciprocal links from other users. This study shares this point of view. According to Digital News Report 2021, a comparative analysis of media usage in 46 countries shows that television continues to be a powerful medium in Japan~\cite{Newman:2021}. According to this investigation, when asked what media they used as news sources, 63\% of Japanese respondents answered that they used online media (including social media) and 58\% chose television.

\section{Data and Methods}
This study draws on two datasets: Twitter data collected by the authors, and television-related metadata compiled by a Japanese company, ``M Data Co., Ltd.'' specializing in collecting metadata related to television. For the Twitter data, 7.1 million cases were collected between December 1, 2020, and June 22, 2021, with the incorrect term ``変異種 (strain)'' and the correct terms ``変異株 (variant)'' or ``変異ウイルス (COVID-19 variant).'' In some analyses, 5,000 or 10,000 cases, not all data, were randomly extracted and analyzed. Television data were acquired through a system called ``i-Catch'' provided by VLe Linac, Inc. The system provides headlines and notes for news items aired on the news programs of Japan's six dominant broadcasting stations. Through this system, we extracted information about 7,600 news items broadcast from December 1, 2020, to June 9, 2021, that contained the term ``変異'' in headlines or memos. 

\section{Results}
\subsection{Volume of data}
\begin{figure}[tp]
    \centering
    \subcaptionbox{Twitter}
    {
        \includegraphics[width=.94\columnwidth]{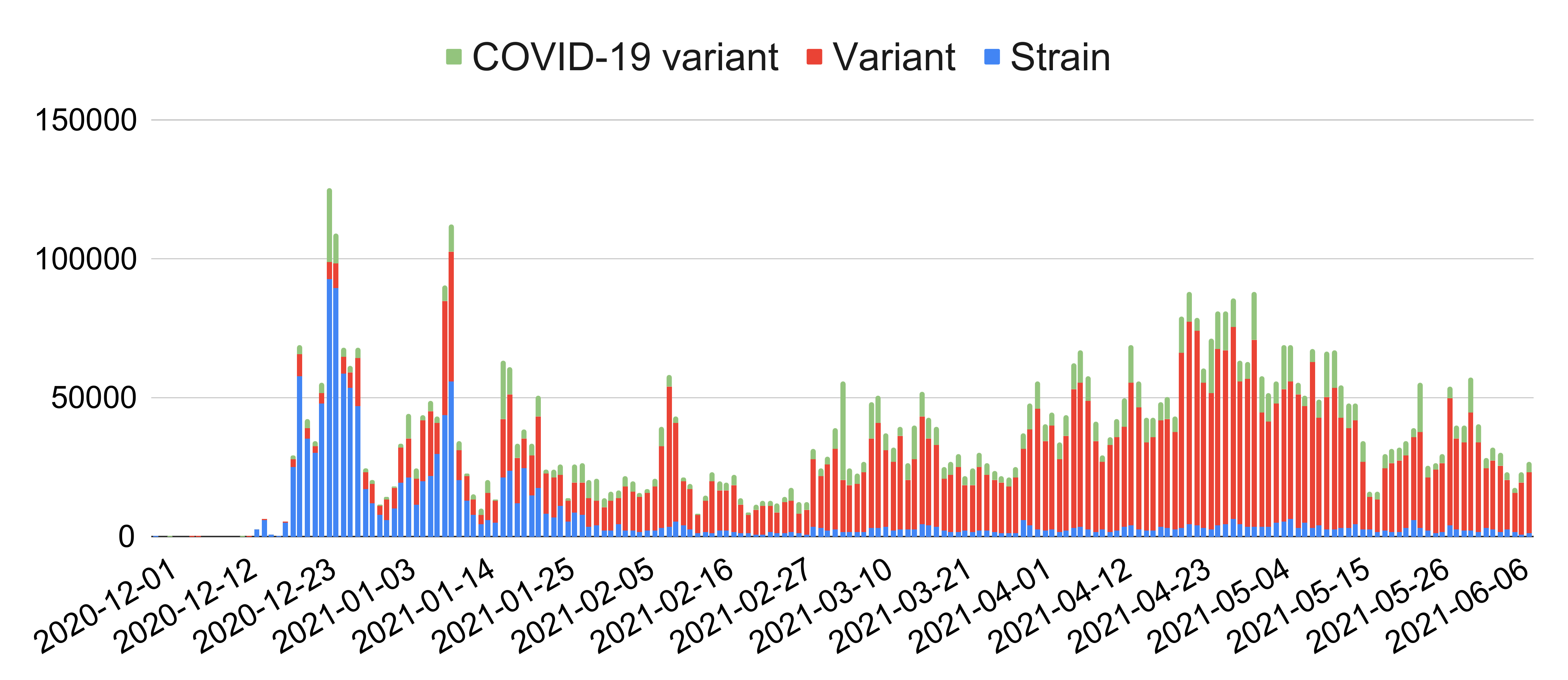}
    }
    \subcaptionbox{Television news programs}
    {
        \includegraphics[width=.94\columnwidth]{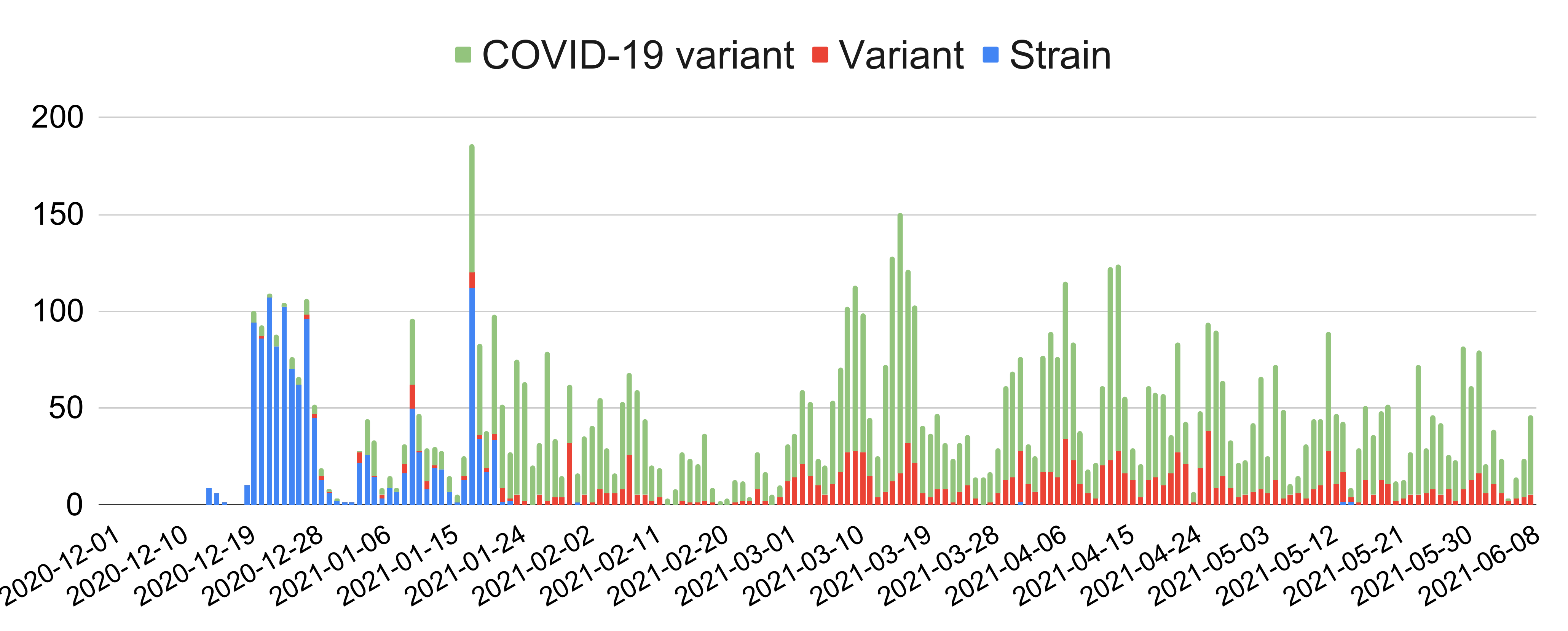}
    }
    \caption{Comparison of the number of incorrect terms (blue) and correct terms (red and green).}
    \label{fig:Comparison}
\end{figure}

\begin{figure}[tp]
    \centering
    \subcaptionbox{Twitter}
    {
        \includegraphics[width=.94\columnwidth]{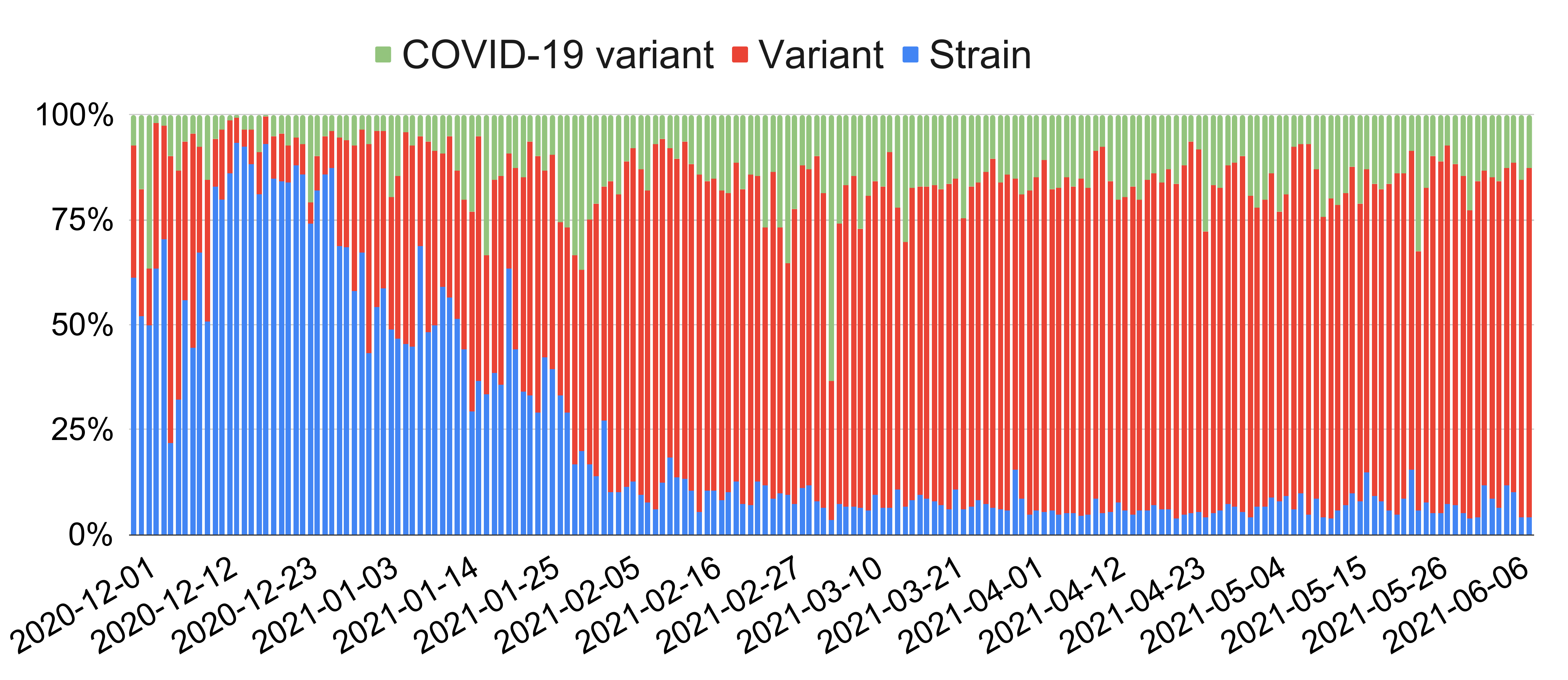}
    }
    \subcaptionbox{Television news programs}
    {
        \includegraphics[width=.94\columnwidth]{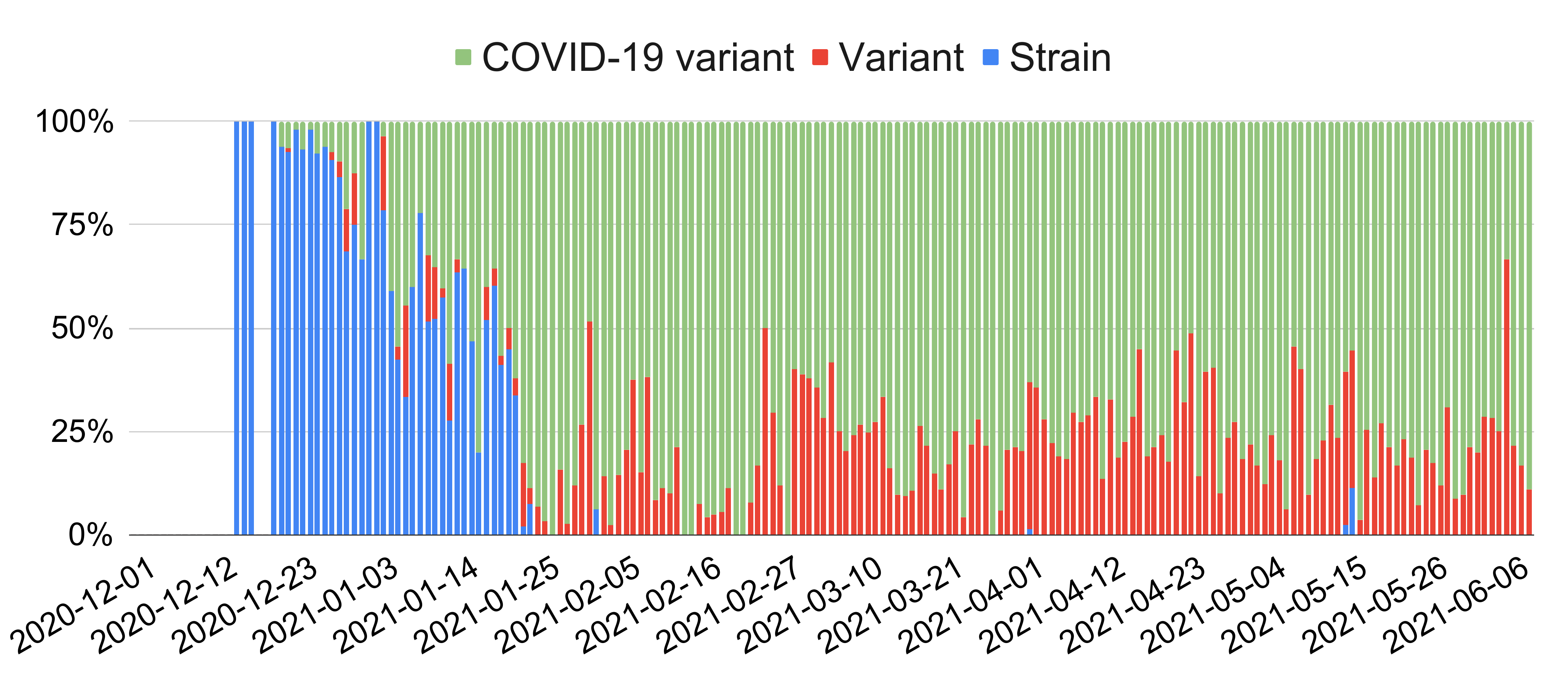}
    }
    \caption{Relative proportion of the number of incorrect terms (blue) and correct terms (red and green).}
    \label{fig:Relative}
\end{figure}

First, we measured the number of tweets on Twitter, including incorrect and correct terms, and visualized them as graphs (Figure~\ref{fig:Comparison}-a). As mentioned above, the incorrect term refers to ``strain,'' represented in blue in the graph, while the correct terms include ``variant'' and ``COVID-19 variant,'' represented in red and green on the graph, respectively. Tweets that used both correct and incorrect terms accounted for only 2\% of the total data; thus, no separate processing was performed, such as deletion. Figure~\ref{fig:Comparison} shows that the number of users using the incorrect term on Twitter (blue) has dwindled since mid-December 2020. Although it increased again in mid-January 2021, we can see that there are relatively more users (red and green) who use the correct term. We performed the same task on television data (Figure~\ref{fig:Comparison}-b). Furthermore, for the same data, we visualized it with a 100\% stacked bar graph to verify the relative ratio. Figure~\ref{fig:Relative}-a shows the analysis results of Twitter data, and Figure~\ref{fig:Relative}-b shows the television data results. 

According to Figure~\ref{fig:Relative}-a, in the case of Twitter, the proportion of incorrect terms started to decrease around December 25, 2020. Specifically, it recorded an 80--90\% decrease in mid-December 2020, 74\% on December 25, 2020 and 68\% from December 29--30, 2020. Figure~\ref{fig:Relative}-b shows the overall similarity in the case of television, but the proportion of incorrect terms, which accounted for a 90--95\% decrease from mid-December, declined throughout January 2021 after hitting 68\% on December 30, 2020. In summary, although the decreasing speed is not constant and it is difficult to clearly measure the speed, given the overall tendency, Twitter registered a relatively low rate of usage of the wrong terms in December 2020 compared to television, and the rate at which it was replaced by the correct term was relatively faster. The proportion of incorrect terms has decreased dramatically since January 2021. Interestingly, after February 2021, while we could hardly find an incorrect term on television, there was a group of users who consistently used the misnomer on Twitter.
\subsection{Retweet ranking}
As demonstrated above, the behavior of Twitter users changed faster to a certain extent than television users. Therefore, what is the information that influences them to change their behavior? To understand this, we analyzed the retweeting behavior of users. We set the analysis period as of December 25--31, 2020 as demonstrated in 4.1, during which period the rate of misuse of the term began to decrease significantly. Table~\ref{tb:TopRetweeted} shows a list of the 10 most retweeted tweets during this period and the users (accounts) who sent those tweets. 

\begin{table*}[tp]
  \centering
  \small
\caption{Top 10 most retweeted tweets from December 25--31, 2020.}
\label{tb:TopRetweeted}
\begin{tabular}{llll}
\toprule
\textbf{rank} & \textbf{tweet (translated to English)}                                      & \textbf{account} & \textbf{remarks}       \\
\midrule
1             & {[}Breaking News{]} People infected with a strain (...)                     & \texttt{@tbs\_news}        & traditional media (TV) \\
2             & That's right. It's a ``variant'', not a ``strain''(...)                        & \texttt{@YamabukiOrca}     & influencer (doctor)    \\
3             & {[}Breaking News JUST IN{]} COVID-19 variant (...)                          & \texttt{@nhk\_news}        & traditional media (TV) \\
4 & The word ``strain'' is frequently   used, but it is not a new ``species'' (...)               & \texttt{@masahirono} & influencer (doctor)     \\
5 & {[}Breaking News{]} Infection by a strain spreads   in Britain and other countries (...)  & \texttt{@tbs\_news}  & traditional media (TV) \\
6             & {[}Breaking News{]} First confirmation in Japan of   COVID-19 variant (...) & \texttt{@livedoornews}     & portal site            \\
7             & The situation regarding a strain of COVID-19   is dizzying (...)            & \texttt{@kutsunasatoshi}   & influencer (doctor)    \\
8 & Confirmed infection of a man in his thirties   who returned from the United Kingdom (...) & \texttt{@ojimakohei} & influencer (politician) \\
9             & Locked down in the UK, some areas work and   some don't (...)               & \texttt{@shinkai35}        & influencer             \\
10            & {[}Asahi Shimbun extra edition{]} Extra: Strain of   COVID-19 (...)         & \texttt{@UN\_NERV}         & influencer            \\
    \bottomrule
\end{tabular}
\end{table*}

According to the analysis results, the accounts ranked at the top can be categorized into 1) traditional media, 2) portal sites, and 3) influencers, including doctors and politicians. Accounts operated by television and portal sites, which have a strong influence on Japan's media environment, also had a significant impact on Twitter. Influencers, particularly doctors, were ranked at the top. This could be attributed to the characteristics of the analysis target, which is terminology related to COVID-19. Based on these results, we predicted that there could be a difference in behavior among users who retweeted information from each of these three groups (traditional media, portal sites, and influencers). To verify this, we analyzed whether there were any changes in behavior before and after the analysis period set in Table~\ref{tb:TopRetweeted} (December 25--31, 2020, hereinafter ``period B''). Specifically, we investigated the percentage of incorrect terminology used in the week before (December 18--24, 2020, hereinafter ``period A'') and the week after (January 1--7, 2021, hereinafter ``period C''). Based on the Retweet Top 10 analysis results (Table~\ref{tb:TopRetweeted}), accounts of the top two traditional media (\texttt{@tbs\_news} and \texttt{@nhk\_news}), the top two portal site (\texttt{@livedoornews} and \texttt{@YahooNewsTopic}), and the top two influencers (\texttt{@YamabukiOrca} and \texttt{@masahirono}) were selected as representatives for each group. Although not included in the table, \texttt{@YahooNewsTopic}, the most influential portal site in Japan, ranked 12th. 

\begin{figure}[tp]
    \centering
    \subcaptionbox{All Twitter users}
    {
        \includegraphics[width=.92\columnwidth]{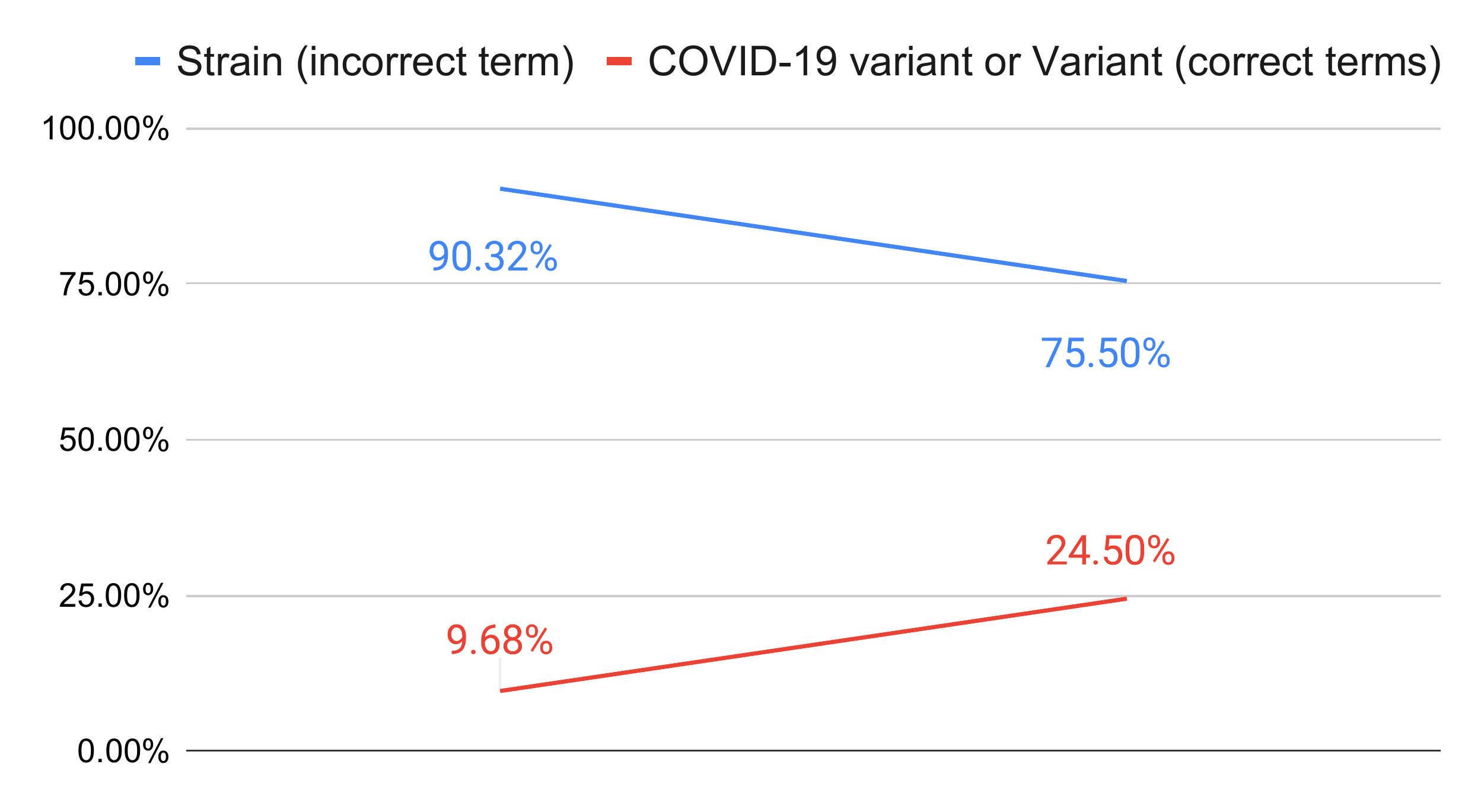}
    }
    \subcaptionbox{Users who retweeted tweets from traditional media}
    {
        \includegraphics[width=.92\columnwidth]{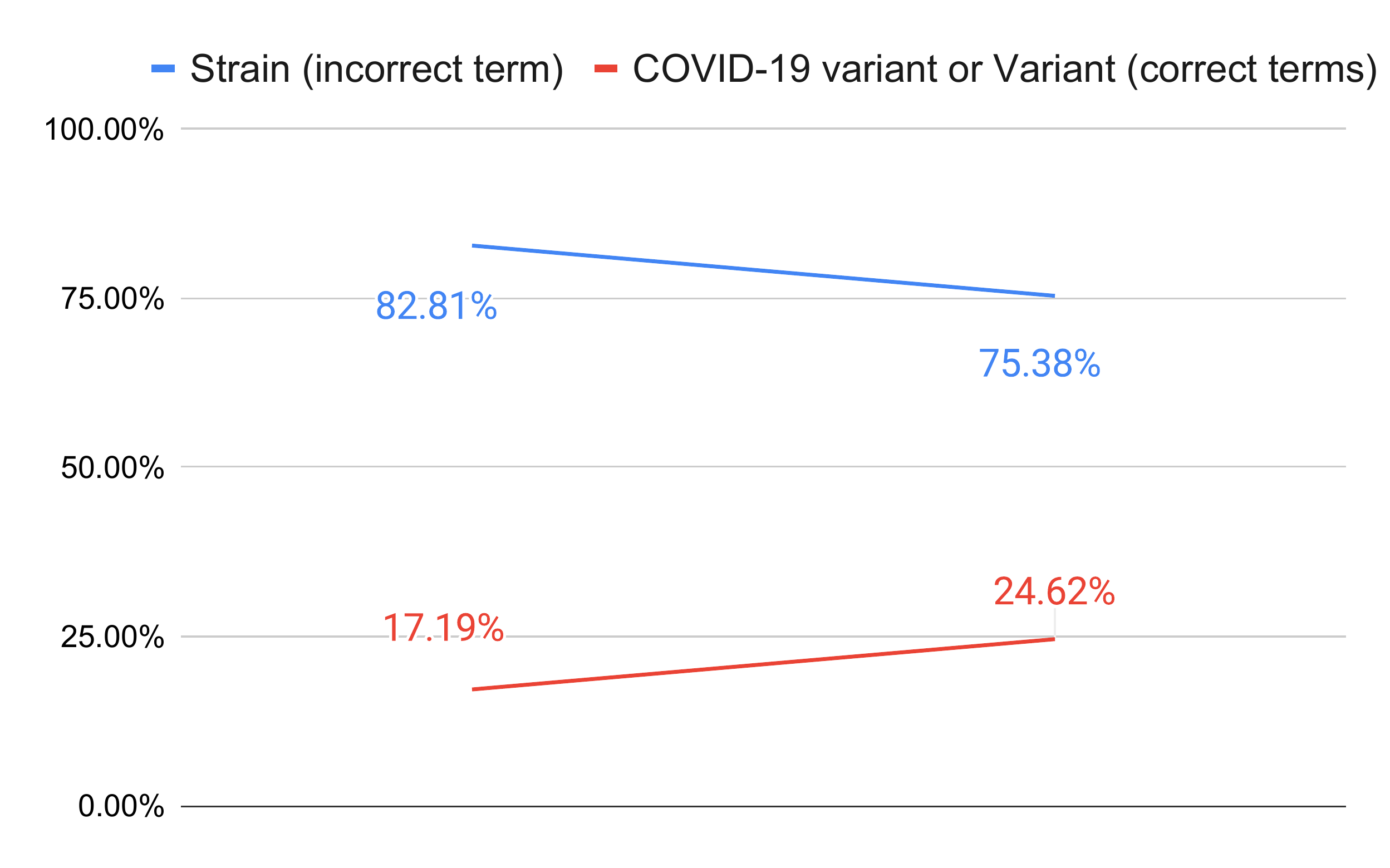}
    }
    \centering
    \subcaptionbox{Users who retweeted tweets from portal sites}
    {
        \includegraphics[width=.92\columnwidth]{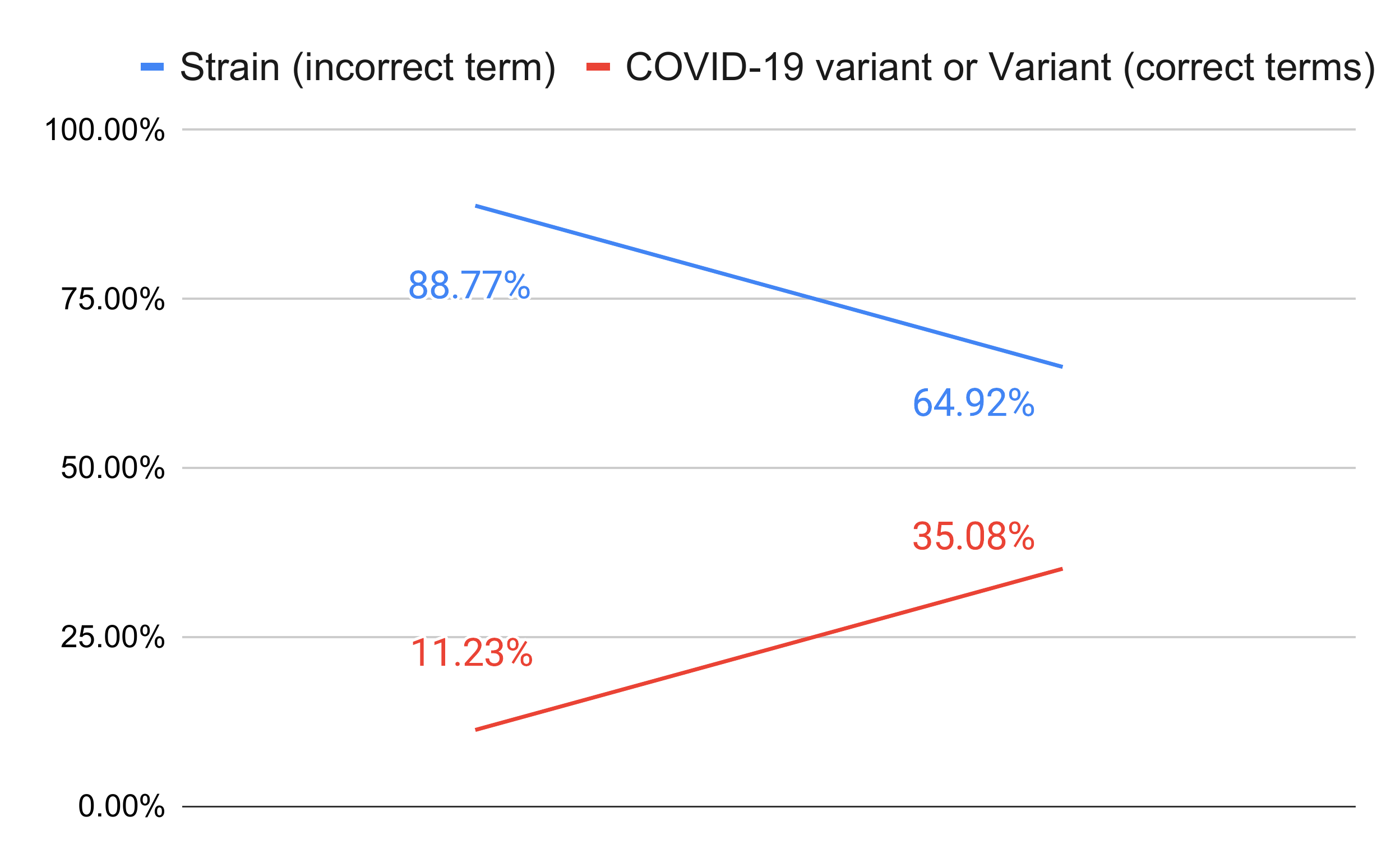}
    }
    \centering
    \subcaptionbox{Users who retweeted tweets from influencers}
    {
        \includegraphics[width=.92\columnwidth]{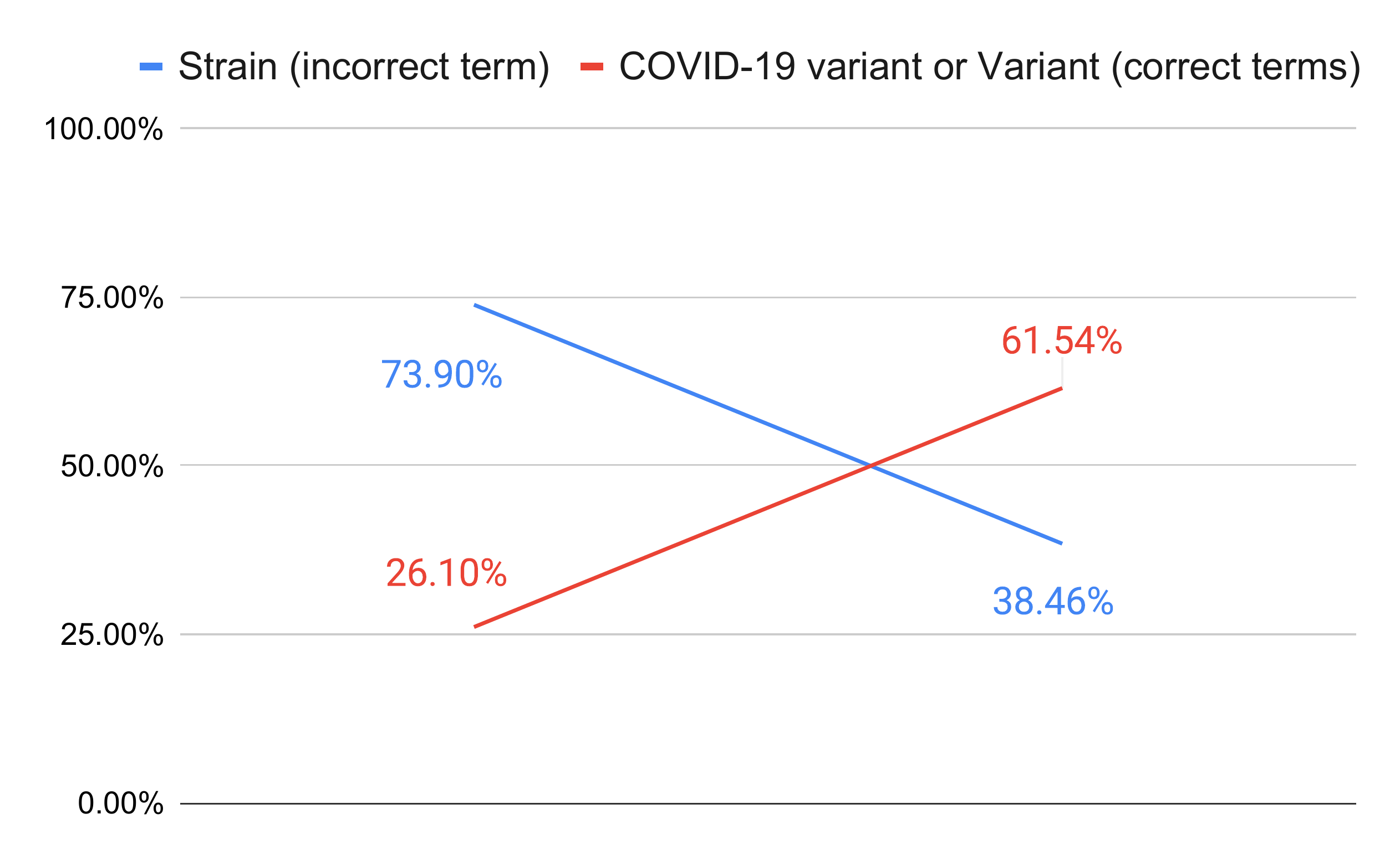}
    }
    \caption{Changes in the percentage of incorrect terminology.}
    \label{fig:Changes}
\end{figure}

Figure~\ref{fig:Changes}-a is a control group for comparison, which is the result of an analysis of all users regardless of the retweeted media. Although the proportion of users using the correct term increased to 24.5\% in period C, 75.5\% continued to use the incorrect term. Figure~\ref{fig:Changes}-b shows that, in the case of users who retweeted traditional media tweets during period B, 75.3\% used the incorrect term and 24.6\% used the correct terminology during period C. In the group of users who retweeted tweets from the portal sites shown in Figure~\ref{fig:Changes}-c, 64.9\% used the incorrect term and 35\% used the correct expression during period C. Finally, Figure~\ref{fig:Changes}-d shows that for users who retweeted tweets from influencers, the proportion of those who used the correct term in period C increased to 61.5\%. In addition, the percentage of users who used the incorrect term was 38.4\%. This was the only case in which the correct terms were used more than 50\% of the time.

\subsection{People who continued to use the incorrect term}
As confirmed in 4.1., which analyzed changes in volume, the proportion of users using the wrong term on Twitter has declined sharply since January and February 2021, they have become a minority. However, even after March 2021, when the correct terminology was widely used, around 5\% of users consistently used the incorrect term. We wanted to learn more about those who ``resist change'' or ``still have not accepted new information.'' We extracted and analyzed URLs inserted in the form of hyperlinks in the body of the tweets they sent or retweeted. We expected that this analysis would allow us to learn what media they are referring to other than Twitter. In addition, owing to the limitations of the analysis tool, 5,000 cases were randomly extracted and analyzed from each set.

Table~\ref{tb:TopSitesCorrect} shows the analysis results of the groups using the correct term after March 2021. Among the accounts corresponding to the Top 10, eight are traditional media; one is a portal site; and YouTube holds the fifth place. Table~\ref{tb:TopSitesIncorrect} shows the media mainly cited by users who continued to use the incorrect term after March 2021; it is worth noting that YouTube is ranked first. It is also interesting that sports newspapers are in the eighth and ninth places, and a conspiracy theory-related site is in tenth place. 

\begin{table}[tp]
  \centering
  \caption{Top 10 sites cited by Twitter users who continued to use the correct term after March 2021.}
  \label{tb:TopSitesCorrect}
    \begin{tabular}{llr}
        \toprule
    \textbf{rank} & \textbf{sites}        & \textbf{cited} \\
        \midrule
    1             & news.yahoo.co.jp      & 173            \\
    2             & mainichi.jp           & 92             \\
    3             & www.tokyo-np.co.jp    & 83             \\
    4             & www.jiji.com          & 71             \\
    5             & www.youtube.com       & 69             \\
    6             & nordot.app            & 65             \\
    7             & www.nikkei.com        & 62             \\
    8             & jp.reuters.com        & 51             \\
    9             & www.newsweekjapan.jp  & 44             \\
    10            & www.nikkan-gendai.com & 36             \\
    \bottomrule
    \end{tabular}
\end{table}

\begin{table}[tp]
\centering
\caption{Top 10 sites cited by Twitter users who continued to use the incorrect term after March 2021.}
\label{tb:TopSitesIncorrect}
\begin{tabular}{llr}
    \toprule
\textbf{rank} & \textbf{sites}         & \textbf{cited} \\
    \midrule
1                & www.youtube.com        & 172            \\
2                & news.yahoo.co.jp       & 55             \\
3                & www.nikkei.com         & 39             \\
4                & ameblo.jp              & 31             \\
5                & www.abc.net.au         & 31             \\
6                & www.yomiuri.co.jp      & 29             \\
7                & jp.reuters.com         & 28             \\
8                & www.tokyo-sports.co.jp & 28             \\
9                & www.nikkansports.com   & 25             \\
10               & indeep.jp              & 20             \\
\bottomrule
\end{tabular}
\end{table}

\begin{figure}[tp]
    \centering
    \subcaptionbox{All tweets including URL}
    {
        \includegraphics[width=.94\columnwidth]{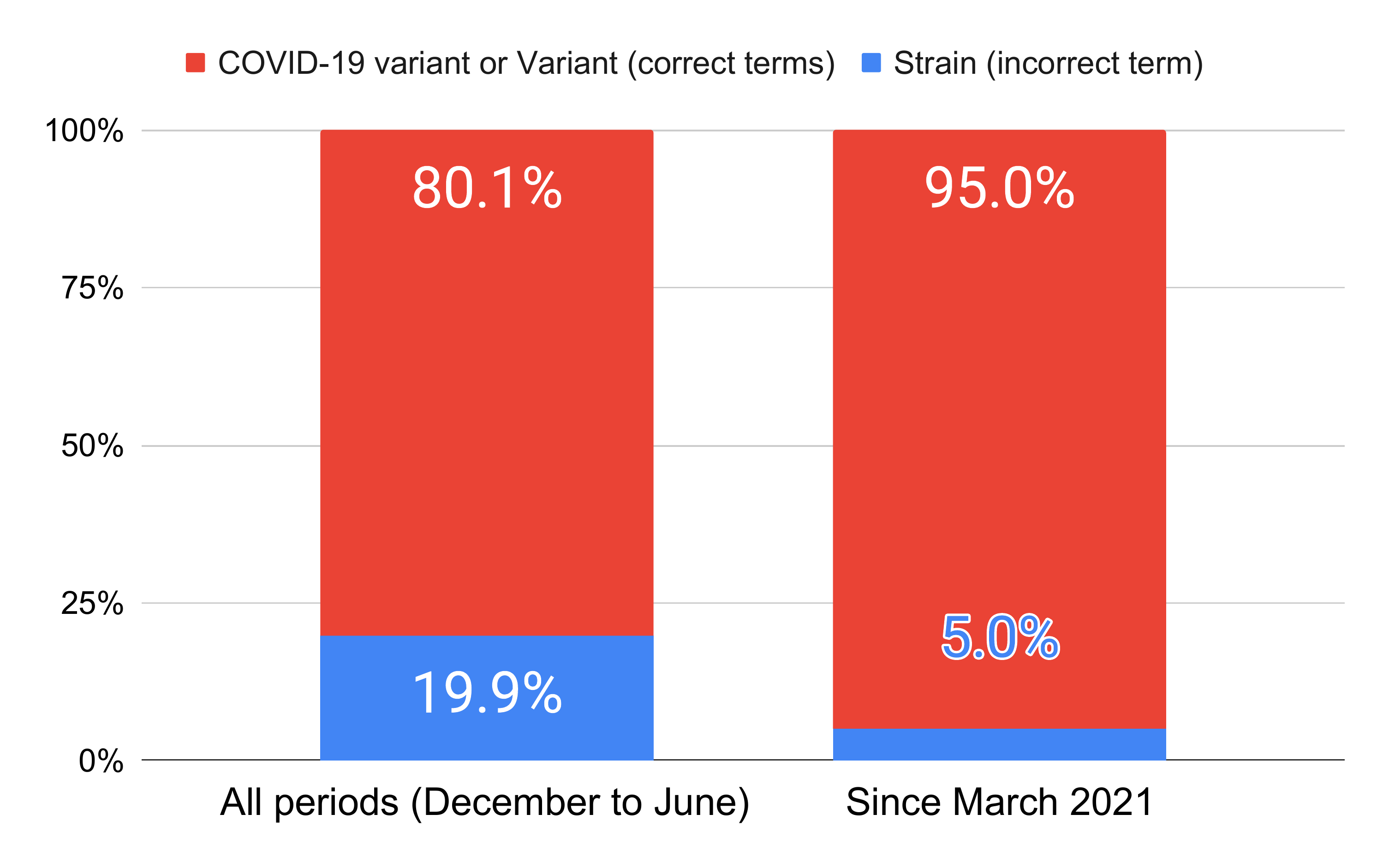}
    }
    \subcaptionbox{Only tweets including YouTube URL}
    {
        \includegraphics[width=.94\columnwidth]{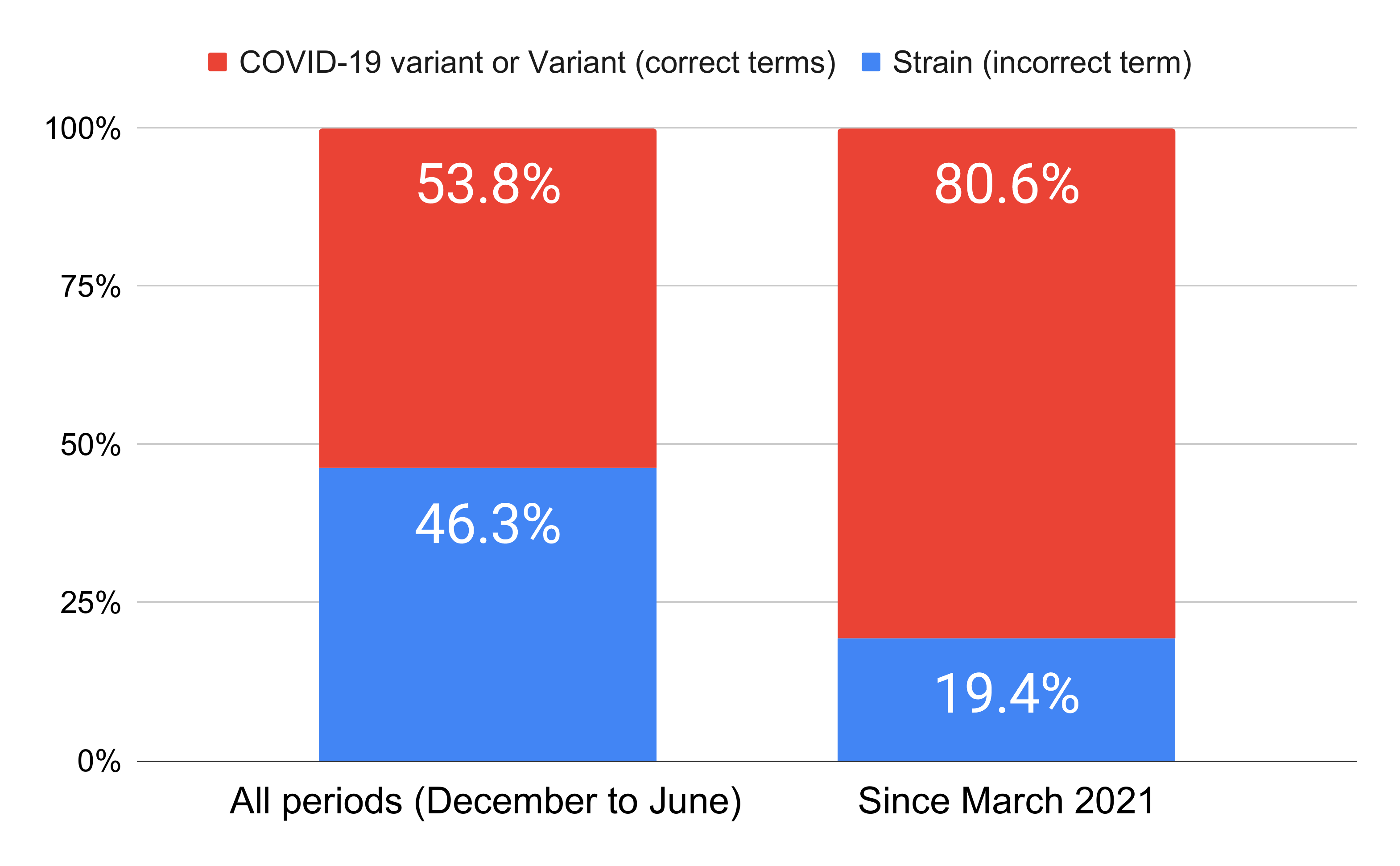}
    }
    \caption{Percentage of incorrect terms.}
    \label{fig:Percentage}
\end{figure}

In addition, based on the results shown in Table~\ref{tb:TopSitesCorrect} and Table~\ref{tb:TopSitesIncorrect}, we analyzed whether there is a difference in the percentage of correct term usage depending on the referenced URL, that is, the referenced media. In this analysis, 10,000 cases were randomly extracted and set as the target for analysis. Figure~\ref{fig:Percentage}-a shows the analysis of the control group, which targeted all tweets with URLs attached in the form of hyperlinks. When analyzing tweets over all periods (December 2020 to June 2021), 19.9\% of tweets used the wrong terminology, and after March 2021, it fell to 5\%. However, the analysis results of the tweets citing YouTube are shown in Figure~\ref{fig:Percentage}-b. It showed 46.3\% for all periods and 19.4\% since March 2021, indicating that there are far more users who use the incorrect term than the control group.

\section{Discussion and Conclusions}

In this study, we analyzed how scientifically-correct information is disseminated on Twitter. We determined the process of correct terms replacing wrong expressions by analyzing 7.1 million Twitter data. We also analyzed more than 7,600 television metadata for comparison. Accordingly, our answers to each of the three main points presented in the introduction are as follows:
\begin{itemize}
  \item The rate of the use of a scientifically-incorrect term ``strain'' on Twitter began to decline in late December 2020 and continued to decline in January 2021, with more than 90\% being replaced by the correct terms in February 2021. The changes began to occur slightly earlier than they did for television.
  \item When users were grouped according to the account from which they retweeted, the rate at which they started using the correct term differed. Users who retweeted tweets from traditional media and portal sites began to use the correct term faster than the overall average, but more than half continued to use the incorrect term. In contrast, for users who retweeted influencers on Twitter, the correct term usage rate was 66\%, exceeding the incorrect term usage rate of 33\%.
  \item Even after March 2021, when most users were using the correct term, some users continued to tweet the wrong term; YouTube was the most cited media by such users. This was markedly different from other groups, which tended to mainly cite portal sites and traditional media. In addition, when we extracted tweets citing YouTube and analyzed the percentage of incorrect terms used, it was higher than that shown in tweets citing the portal site Yahoo! JAPAN.
\end{itemize}

Regarding the first result, we assume that, to some extent, self-corrections have occurred on Twitter. This is because the incorrect term was already being rectified before January 22, 2021 when the Japanese Association for Infectious Diseases officially issued a statement on the correct term. It was even a little ahead of television, a major media outlet with a strong influence in Japan. Furthermore, this was a case in contrast to the failure to correct misinformation, such as the toilet paper rumor in Japan in March 2020. This can be attributed to the fact that users were not strongly opposed to the corrected information because the subject of this analysis was a matter of terminology that had nothing to do with individual beliefs. The fact that users who continued to use the wrong term after March 2021 are still active on Twitter reportedly highlights the limitations of self-correction on social media.

In the second result, the fact that three of the top five accounts that sent the most retweeted tweets related to traditional media (especially television) and portal sites seems to reflect Japan's unique media environment. Given that portal sites generally serve as a conduit for the collection and provision of articles originating from traditional media, the influence of traditional media, especially television, is quite strong.

Among the accounts of influencers on Twitter, which accounted for the remaining 6 of the top 10 list, 3 accounts indicated that they were doctors. This seems to be related to the fact that the subject of this study was a medical term related to COVID-19. It makes sense to trust a medical expert's message regarding medical jargon. However, it is also worth noting that there was a difference in behavior between users who retweeted tweets from traditional media or portal sites and those who retweeted tweets from influencers. This could be attributed to differences in the content of the message. Specifically, tweets from traditional media or portal sites were mainly breaking news, while some influencers graciously explained the correct use of terminology, such as ``it is right to use the term variant rather than strain.'' Specifically, tweets by influencers containing the correct term ranked second and fourth in the top 10 list. This period was immediately after the outbreak of the variant, and there was still no clear guidance from authorities on the specific terminology. Therefore, it can be inferred that influencers with expertise may have had a significant impact on Twitter users' use of jargon.

The third result is an intensive analysis of users who resisted change and continued to use the incorrect term, even after the correct terminology was widely prevalent. Admittedly, we have not been able to determine whether they are deliberately using the incorrect term, or whether they continue to use it due to a lack of information. However, through URL analysis, it was possible to determine what media they primarily cited. Compared with the results of analyzing the control group and users using the correct term, users of the incorrect term cited YouTube more frequently. In addition, the fact that sites mainly dealing with conspiracy theories were ranked in the 10th place was also a characteristic phenomenon not found in other groups. It was also found that the proportion containing incorrect terms was significantly higher than those containing other URLs. It can be inferred that there may be a relationship between users who cite YouTube and those who use incorrect terminology. 

This study has several limitations. First, one might question the external validity of this research in that it is a case study. Furthermore, since it is a special case in which influencers, experts on terminology, have had a significant impact and have succeeded in rectifying misinformation, it is highly likely that the same result will not be obtained in other cases. This study analyzed two types of data, Twitter and television metadata, but it is regrettable that better insights could have been obtained if Internet news and newspaper articles, which wield considerable influence in Japan, were also included in the analysis. 

Nevertheless, this study empirically confirms the process by which incorrect information on Twitter can be successfully rectified through interaction within Twitter. This also allowed us to understand the influence of experts on the dissemination of scientific facts. Even on Twitter, which has been recognized as a ``hotbed of rumors and fake news propagation,'' we could see that self-corrections centered on influencers with expertise occurred. Furthermore, we confirmed that users who use the correct term and those who persistently use incorrect terminology differ in the media they cite. This suggests that the type of media one refers to can also affect individual attitude changes.

\begin{acks}
The television news program data used for our study were provided by M Data Co., Ltd. through i-Catch provided by VLe Linac, Inc.
This work was supported by JST RISTEX Grant Number JPMJRX20J3, and JST-Mirai Program Grant Number JPMJMI20B4, Japan.
\end{acks}

\bibliographystyle{ACM-Reference-Format}
\bibliography{main}

%%\appendix

\end{document}